\documentstyle[multicol,aps,prl]{revtex}
\begin{document}
\draft
\tightenlines
\title{Driven diffusive system: A study on large n limit}
\author{Sutapa Mukherji}
\address{Department of Physics and Meteorology, Indian 
Institute of Technology, Kharagpur 721 302, India}
\onecolumn
\date{\today}
\maketitle
\widetext
\begin{abstract}
We study the generalized $n$ component model of a driven diffusive 
system with annealed random drive in the large $n$ limit. This 
non-equilibrium model  also describes conserved  order parameter 
dynamics of an equilibrium model of ferromagnets with dipolar 
interaction. 
In this limit, at zero 
temperature a saddle point approximation becomes exact. The length
scale in the direction transverse to the driving field acquires an 
additional logarithmic correction in this limit.   
\end{abstract}
\begin{multicols}{2}

\section{Introduction}
In nature,  a large class of phenomena are far-from-equilibrium in origin.
Unlike the equilibrium systems, depending on the complexity and the relevant
questions involved, there are different ways to treat  non-equilibrium 
systems. In equilibrium, due to the random energy imparted by  
the environment (thermal noise),
 all possible energy states are accessible
 and there is no net flow of probability (i.e. no net current)
 from one configuration to the
other. As a result there is a straight forward way of obtaining the 
free energy and hence the thermodynamic properties by just summing over all 
the Boltzmann factors $\exp[-E/K_BT]$ for all the states, where $E$ is
 the energy of the state.  In general, although one may write down a 
master equation describing the time evolution of the probability 
of a system being in a state, in the 
large time limit, the equilibrium dynamics ensures 
a Boltzmann 
 distribution. 
This behavior is preserved by the so called detailed balance principle. 
This shows up also in the 
Langevin equation which describes the time evolution of the degrees 
of freedom (say e.g. the order parameter in a spin system, or the height 
of a growing surface). In the equilibrium case, the Langevin equation 
can be obtained from the Hamiltonian representing all the relevant 
interactions of the system. However, in the non-equilibrium cases, such a
description through a Hamiltonian is not possible due to one or more 
terms.  These terms are purely of non-equilibrium origin and  appear
due to, say, a finite driving force in the driven dynamics of particles
\cite{ziarev}
 or in random deposition of particles in the case of crystal growth 
\cite{halpin}. There are  non-equilibrium systems where detailed 
balance may also hold good. In these cases the dynamics reduces to the 
equilibrium dynamics of a system described with appropriate 
Hamiltonian \cite{halpin}.

These aspects of non-equilibrium dynamics are  
present in different variants of driven diffusive systems (DDS) 
\cite{ziarev} 
where particles with diffusive dynamics have biased hopping due to a 
driving force in a particular direction.
In the case of fixed time independent driving force, a formulation 
of the Langevin equation through a Hamiltonian is not possible due to the 
term responsible for the drive.  This is not necessarily true for all 
the driven systems. In particular, we shall later be concerned with a 
driven system with an annealed random drive where the problem effectively 
reduces to the equilibrium dynamics of a specific magnetic system.
In general, without the driving force, the equilibrium system 
is like a lattice gas where particles
randomly occupy lattice sites and also due to a nearest neighbor attraction
a particle  prefers to have  its next neighboring site occupied by another 
particle. In this sense the equilibrium static 
properties are the same as that 
of the Ising model with a ferromagnetic interaction. Even though, the dynamics,
 unlike the spin case, is  different due to the total particle
number conservation in the lattice gas, the static properties of the two 
models are the same due to the equivalence of  canonical and 
grand-canonical ensembles.
With a finite driving force \cite{kls}, the dynamics of  
lattice gas is significantly 
different from the equilibrium order parameter conserving dynamics.
 Three key features, such as, the   
boundary condition, particle number 
conservation and the driving force together  affect 
the dynamics in a nontrivial way and as a result one observes
generic power law correlation far above the criticality and  anisotropy in 
the exponents of different correlation functions \cite{ziarev}. 

Due to the time dependent probability distribution, and lack of detailed 
balance,  many questions remain unresolved for this system.  Although in 
the steady state, where the distribution settles to a time independent 
form, one can write down a local detailed balance form for the transition 
rates in the Master equation, obtaining the physical properties from this
is, in practice, difficult. However, even working with very small lattices,
 certain interesting features can  be observed \cite{zhang,katz}. 
For example, on 
a small size half filled lattice, in the ground state, unlike the 
equilibrium case, one can see a strip-like occupied region in the presence 
of a large driving force. A careful analysis of the internal energy 
fluctuation  and specific heat also reveals new consequences 
of the driving force \cite{ziarev}.

In view of the complications, it is, therefore, 
often useful to adopt different approximation schemes or numerical
techniques. Since in many cases the interest is mostly on 
long wavelength
and long time features, the continuum formulation in terms of the Langevin 
equation comes quite handy. 
The continuum formulation  provides a fruitful basis
for performing the mean field type calculations or field theoretic 
renormalization group analysis \cite{janrg}
with the latter being especially successful 
in explaining the new features at the critical point. In this scheme, the 
existence of a non-equilibrium stable fixed point that governs the new 
universal properties, anisotropic non-Gaussian scaling exponents associated 
with it, is clearly observed.
The fact that far above the criticality,  the two point  
or three point correlation exhibits a power law behavior, can be 
directly shown using the above continuum formulation and  
averaging  with $\exp[-{\cal J}]$ with ${\cal J}$ as the dynamic functional. 
This weight factor essentially plays a role that is 
 analogous to the Boltzmann weight for the equilibrium system.

Although significant progress has been made in understanding the 
properties far above and at criticality, the situation below  criticality   
is relatively unclear. Below the critical temperature, one would expect 
 phase separated occupied and unoccupied regions provided the filling 
is not too far away from the half filling where one would remain out of the 
coexistence regime. In the presence of the driving force, the shape of the 
coexistence curve changes from the equilibrium situation 
\cite{vallesltemp} and also the shape 
of the phase separated regions gets distorted. The basic questions that 
 still remain are  related to these aspects. Although numerical 
simulations show that the particle occupied regions are somewhat elongated 
in the direction of the field, sufficient analytical progress is yet to be 
made.      

Along with this development, several variants of this simple case of 
driven lattice gas have appeared in order to understand more realistic 
phenomena in nature. One simple choice of the boundary condition is 
the periodic one in both transverse and longitudinal direction.
In that case the system looks like a torus with the driving field, 
possibly  a uniform  electric field looping around it. 
To obtain such a situation in laboratory, one  would require
a magnetic flux increasing linearly in time. 
Therefore due to  these practical purposes, 
a demand for more abundant situations 
with non-uniform drive or different 
boundary conditions was obvious. Thus 
came up a simpler but more realistic model of DDS 
with annealed random drive
\cite{schmit1,schmit2}.
Interestingly, this model exhibits new universal properties 
that are different from 
the uniformly driven Ising model.

In this paper, we are interested in studying a driven diffusive system 
with annealed random drive with $n$ component order parameter in the large 
$n$ limit. In this limit the problem is exactly solvable at zero 
temperature. It is known that the large time behavior of a 
system, quenched below the critical temperature, is characterized by the 
equal time structure factor
\begin{eqnarray}
c(k,t)=\langle \phi_i({\bf k},t) \phi_i(-{\bf k},t)\rangle=
L^{\alpha}(t) F(k L(t)),
 \end{eqnarray}
where $L(t)\sim t^{1/z}$, with $z$ being the dynamic exponent, 
represents the 
characteristic size of the correlated regions growing with time
\cite{brayscal}. In the above definition, the subscript $i$ denotes the 
component of the order parameter field $\phi$. This scaling 
law is preserved in the case of quenching of a system with a 
non-conserved 
order parameter. However, this scaling form is not maintained in a 
 system with a conserved order parameter. In the latter case one has two 
length scales, obeying two different  scalings \cite{conig,emott}.
 While  the growth of one length scale with time is governed by the
usual dynamic exponent, for the other length scale, an additional time 
dependence appears.
The reason for such multi-scaling is 
understood to be the non-commutativity 
of the large $n$ and large $t$ limits \cite{hum}.

\section{Model}
In the lattice gas systems with a fixed total number of particles, there
  is an underlying  continuity equation obeyed by the particle density
  $\rho$, $\partial_t \rho=-\nabla \cdot J$, where $J$ is the current.
  In the absence of the driving force, the current can be expressed as
   the gradient of the chemical potential $-\lambda \nabla \mu$, where
   $\lambda$ is the transport coefficient, $\mu=\delta {\cal H}/\delta
		 \rho$, with ${\cal H}$ as the appropriate lattice gas
   Hamiltonian. With an attractive
       interaction among the particles,
the  complete description of the dynamics
 involves only one field, the particle density which plays the role of
 the order parameter. Near equilibrium, the equation of motion for the
  conserved dynamics of such attracting particles can be written using
   the continuity equation  and the Hamiltonian, expressed in terms of
    the local magnetization   $\phi({\bf x},t)=2\rho({\bf x},t)-1$, as
\begin{equation}
   {\cal H}(\phi)=\int d^dx\{\frac{1}{2}(\nabla \phi)^2+\frac{\tau}{2}
					 \phi^2+\frac{u}{4!} \phi^4\}.
\end{equation}
 The drive, which is actually responsible for the far-from-equilibrium
   behavior, gives rise to an additional current, which must vanish if
     there is no hole or no particle. Thus, the simplest choice for the
    current may be $J_E=4 \rho(1-\rho)E=(1-\phi^2)E$, where $E$ is the
   strength of the driving field. The system is anisotropic due to the
  driving force in a particular direction, henceforth
  denoted by   $z$. In  the  equation of motion, this $z$ direction  
and the transverse
 $d-1$ dimensional space  are  distinguished by different coefficients
	\cite{ziarev}				as shown below
\begin{eqnarray}
     \partial_t \phi({\bf x},t)=\lambda\{(\tau_\perp-\nabla^2)\nabla^2
       \phi+(\tau_\parallel-\alpha_\parallel \partial^2)\partial^2\phi
\nonumber\\
-	   \alpha_\times \partial^2 \nabla^2\phi+\frac{u}{3!}(\nabla^2
       \phi^3+\chi \partial^2 \phi^3)+E \partial \phi^2\}-(\nabla\cdot
						  \xi+\partial \zeta),
\end{eqnarray}
 Here and in the following discussion, 
 $\nabla$ in general extends over the $d-1$
  dimensional space and $\partial$ denotes the derivative with respect
    to $z$ only. The last two terms in the Langevin equation originate
	from the usual noisy part of the current and have  short range
       correlations both in space and time with variances $n_\perp$ and
  $n_\parallel$. The coefficients associated with the derivatives with
	respect to transverse or parallel  coordinates  have subscripts
      $\parallel$ and $\perp$ respectively. As the critical temperature is
		approached, the most realistic choice of parameters is
 $\tau_\parallel>0$ and $\tau_\perp\rightarrow 0$ \cite{leu}. In fact,
  it  can be explicitly shown  that while the transverse direction, and
	    hence $\tau_\perp$, remain unaffected by the driving force,
 $\tau_\parallel$ is renormalized upwards due to the driving force and
				 remains uniform for the whole system.

 The focus of the present paper is on  an  annealed random drive which
		       has a correlation  $\langle E({\bf x},t) E({\bf
	 x}',t')\rangle=\sigma\delta({\bf x}-{\bf x}')\delta(t-t')$.  A
well-known starting point for the dynamics is to write down the dynamic
  functional ${\cal J}$  that involves the Martin-Siggia-Rose response
   field $\tilde\phi$ \cite{msr}. 
%It can be found that  after averaging out the drive, the term involving
% the 
%variant of the distribution is irrelevant and therefore can be neglected. 
      After averaging out the drive, and ignoring all the redundant or
	 irrelevant terms, the generating functional  can be written as
\begin{eqnarray}
	&&	  {\cal J}_b=\int  dt d^d x\lambda\{\tilde \phi({\bf
		 x},t)[\lambda^{-1}\partial_t-(\tau-\nabla^2)\nabla^2-
\partial^2]\phi({\bf x},t)-\nonumber\\
&&\frac{u}{3!}\tilde\phi\nabla^2\phi^3+\tilde
			       \phi \nabla^2\tilde\phi\}.\label{aneal}
\end{eqnarray}
 Since the non-Gaussian interaction term has no longitudinal operator,
      it does not affect $\tau_\parallel$, which, as a result, has been
 considered to be a constant (unity) in (\ref{aneal}). The coefficient
of the last term in the curly bracket has also been set to unity by 
suitable reparametrization.  For convenience
	    we have denoted $\tau_\perp$   in this equation by $\tau$.

	   It is now appropriate  to point out that there exists 
a direct connection between the driven system and uniaxial ferromagnets
 with dipolar interaction \cite{schmit1,aha,zinn}. 
The generating
	       functional can be written in the detailed balance form as
\begin{equation}
   {\cal J}_b=\int \{\tilde\phi \partial_t \phi-\tilde\phi\lambda {\nabla^2
    }[\frac{\delta {\cal H}_d}{\delta\phi}-\tilde\phi]\},\label{detail}
\end{equation}		
						where 
\begin{eqnarray}
				     {\cal H}_d[\phi]=\int_k \frac{1}{2}
			   \phi(-k)S_0^{-1}(k)\phi(k)+\nonumber\\
\frac{u}{4!}\int_{k_1,k_2,k_3}
			\phi(k_1)\phi(k_2)\phi(k_3)\phi(-k_1-k_2-k_3),
\label{hamfer}
\end{eqnarray}
				   with $S_0^{-1}(k)=k_\perp^{-2}[\tau
  k_\perp^2+k_\perp^4+k_\parallel^2]$. This Hamiltonian represents the
       static system of uniaxial ferromagnet with dipolar interaction
 \cite{aha,zinn}. Therefore studying this driven system with annealed 
random drive also implies studying the order parameter conserved 
dynamics of uniaxial ferromagnets. 

\section{Results}
Our starting point for the large $n$ analysis is essentially  
an $n$ component generalization of 
the Hamiltonian ${\cal H}_d$ appearing in the detailed balance form above.
To proceed further, we start with the Langevin equation which 
represents the dynamics of the driven diffusive system of our 
interest as well as the conserved dynamics of the uniaxial 
dipolar system
\begin{eqnarray}
\frac{\partial \phi_i}{\partial t}=\nabla_\alpha 
[\nabla_\alpha\frac{\partial {\cal H}}
{\partial \phi_i}+\zeta_\alpha(x,t)],
\end{eqnarray}
where $\zeta_\alpha(x,t)$ represents the short range correlated Gaussian 
noise. Here $\alpha\ (=1,\ 2,...d)$ stands for 
the dimension of the space  and
$i\ (=1,....n)$ denotes the order parameter component. 
%\begin{eqnarray}
%\langle \zeta(x,t)\zeta(x',t')\rangle=\overline \Delta \delta(x-x')
%\delta(t-t').
%\end{eqnarray}
For the present discussion, we shall restrict ourselves to the zero 
temperature situation where we need not bother about the noise term. 
%(It 
%yet to be understood if the temperature is a relevant variable around 
%this special zero temperature situation. If the 
%temperature is an irrelevant variable around this point, the result 
%can be useful for low temperature region.)

In Fourier space the, Langevin equation becomes
\begin{eqnarray}
\frac{\partial \phi_i}{\partial t}=(-\tau k_\perp^2-k_\perp^4- 
k_\parallel^2)\phi_i-\frac{u_0}{3!}a(t) k_\perp^2 \phi_i\label{diff1},
\end{eqnarray}
where $\phi^2=\sum_{j=1}^n\phi_j^2=n\langle \phi_i^2\rangle=a(t)$. Solving
(\ref{diff1}), we obtain 
\begin{eqnarray}
\phi_i(k,t)=\phi_i(k,0) e^{-(\tau k_\perp^2+k_\perp^4+k_\parallel^2)t-
({u_0}/{3!}) k_\perp^2 b(t)},
\end{eqnarray}
with $b(t)=\int_0^t a(t') dt'$. Therefore 
\begin{eqnarray}
&\langle \phi_i(x,t) \phi_i(x,t)\rangle=\int \frac{dk}{(2\pi)^d}
 \frac{\Delta}{n} \times \nonumber\\ & 
e^{-(2\tau k_\perp^2+2 k_\perp^4+2k_\parallel^2)t-({u_0}/{3})k_\perp^2 b(t)},
\end{eqnarray}
where we have used the  initial condition 
\begin{eqnarray}
\langle \phi_i({\bf k},0)\phi_j({\bf k'},0)\rangle
=\frac{\Delta}{n}\delta_{ij}\delta(\bf{k}+\bf{k'}).
\end{eqnarray}
Therefore we have 
\begin{eqnarray}
a(t)=\frac{\Delta}{(2\pi)^d}
\int d^{d-1} k_\perp dk_\parallel e^{-(2\tau k_\perp^2+2k_\perp^4)t
-2k_\parallel^2 t -({u_0}/{3})k_\perp^2 b(t)}\nonumber\\
=\frac{\Delta}{(2\pi)^d} (\frac{\pi}{2t})^{1/2} \frac{2\pi^{(d-1)/2}}
{\Gamma[(d-1)/2]}\int d k_\perp
k_\perp^{d-2} e^{-(2 \tau' k_\perp^2+2k_\perp^4)t},
\end{eqnarray}
where $\tau'=\tau+\frac{u_0}{6}\frac{b(t)}{t}$. Substituting 
$(\frac{1}{\tau'})^{1/2} k_\perp=x$, this equation simplifies to 
\begin{eqnarray}
&& a(t)=\frac{\Delta}{(2\pi)^d}(\frac{\pi}{2t})^{1/2} \frac{2\pi^{(d-1)/2}}
{\Gamma[(d-1)/2]}(\tau+\frac{u}{6}
\frac{b(t)}{t})^{(d-1)/2}\times\nonumber\\ &&\int dx\ x^{d-2} 
 e^{-(x^2+x^4)2 t \tau'^2}.
\label{gendim}
\end{eqnarray}
To solve this equation, we shall consider different dimensions separately.
At $d=2$, the saddle point is at the origin. Therefore we obtain 
\begin{eqnarray}
a(t)^2=B_0(\tau+\frac{u}{6} \frac{b(t)}{t})/t,
\end{eqnarray}
where $B_0={\Delta^2}/{4\pi}$.
Differentiating this equation once, we have 
\begin{eqnarray}
2 t^2 a(t) \frac{d a(t)}{dt}=\frac{u B_0}{6} a(t)+B_0 \tau-2 t a(t)^2.
\end{eqnarray}
The numerical solution of this equation in terms of three 
constants $c_1$, $c_2$ 
and $c_3$ is
\begin{eqnarray}
a(t)=c_1+\frac{c_2}{t^{1/2}}+c_3 \frac{\log t}{t}.
\end{eqnarray}
In the large time limit, the dominant time dependence is of the form 
$a(t)\sim t^{-1/2}$. Finally, since the factor $2 t \tau'^2$ in the 
argument of the exponential approaches infinity as
 $t\rightarrow \infty$, the saddle point approximation turns out to 
be an exact one. 

Let us consider $d=3$. Starting with (\ref{gendim}), we again 
replace the integral by the maximum value of the integral. This occurs at 
\begin{eqnarray}
x^2=\frac{-2c_0\pm\sqrt{4 c_0^2+16c_0}}{8c_0},
\end{eqnarray}
where $c_0=2\tau'^2t$. As $t\rightarrow \infty$, one can approximate 
$x^2\sim {1}/{2c_0}$. Therefore in the large time limit,
\begin{eqnarray}
a(t)=\frac{\Delta}{4(2\pi)^{3/2}} \frac{1}{t}
\exp[-(1/2+1/4c_0)]
\end{eqnarray}
Therefore at $d=3$, $a(t)$ scales as $1/t$ at large time. To obtain 
the relevant length scale, we rewrite $\phi_i(k,t)$ as 
\begin{eqnarray}
\phi_i(k,t)=\phi_i(k,0) e^{-k_\parallel^2 t} 
e^{[(\tau t+({u}/{3!}) b(t))^2/4t]\{1-((k/k_L)^2+1)^2\}}
\end{eqnarray}
Apart from the length scale in the parallel direction growing as 
$L_\parallel \sim t^{1/2}$,
we find that the   length scale in the direction perpendicular
to the driving force acquires a  logarithmic correction 
in addition to the expected scaling behavior. At $d=3$, the length 
scale in the transverse direction scales with time as  
\begin{eqnarray}
L_\perp \sim \frac{t^{1/4}}{(1+u\ln t/6 t^{1/2})^{1/2}}\ \simeq t^{1/4}\ 
{\rm for} \ t\rightarrow \infty ,
\end{eqnarray}
This is somewhat different from the multiscaling observed before, where
there are two different scaling lengths. Here due to the anisotropy 
originally present in the system, the transverse and longitudinal length 
scales $L_\perp$ and $L_\parallel$ respectively, have different scaling 
properties with time even at the Gaussian level. $L_\perp$, due to the 
non-Gaussian interaction, relevant only in the transverse direction, has 
modified scaling behavior due to a logarithmic correction in addition
to the scaling expected at the  Gaussian level. As expected, the scaling
of $L_\parallel$ remains unaltered from the Gaussian level.

To summarize, we have studied the large $n$ limit of a generalized 
$n$ component driven diffusive model with annealed randomness at zero 
temperature. This model also 
corresponds to the order parameter conserved dynamics of 
uniaxial ferromagnets with 
dipolar interaction. This model is exactly soluble in the large $n$ limit.
The scaling of the length scale transverse to the drive 
is only modified due to an additional logarithmic correction. This is quite 
different from the multiscaling behavior observed before in conserved 
dynamical models.

I thank H. W. Diehl for the  hospitality  
at Universit\"at Gesamthochschule Essen, where part of this work was 
performed.

\end{multicols}

\begin{references}
\bibitem{ziarev} B. Schmittmann
 and R. K. P. Zia in {\it Phase transitions and critical 
phenomena} edited by C. Domb and J. L. Lebowitz (Academic, London, 1995), 
Vol XVII
\bibitem{halpin} T. Halpin-Healy and Y. C. Zhang, Phys. Rep. {\bf 254}, 215
(1995); S. Mukherji and S. M. Bhattacharjee, Current Science {\bf 77},
394 (1999)
\bibitem{kls} S. Katz, J. L. Lebowitz and H. Spohn, Phys. Rev. B {\bf 28},
1655 (1983)
\bibitem{zhang} M. Q. Zhang, Phys. Rev. A {\bf 35}, 2266 (1987)
\bibitem{katz} S. Katz, J. L. Lebowitz and H. Spohn, J. Stat. Phys. 
{\bf 34}, 497 (1984)
\bibitem{janrg} 
H. K. Janssen and B. Schmittmann Z. Phys. B {\bf 64}, 503 (1986)
\bibitem{vallesltemp}
J. L. Valles and J. Marro, J. Stat. Phys. {\bf 49} 89 (1987);
J. Marro and J. L. Valles, J. Stat. Phys. {\bf 49}, 121 (1987)
\bibitem{schmit1} B. Schmittmann, Euro. Phys. Lett. {\bf 24}, 109 (1993)
\bibitem{schmit2}  B. Schmittmann and R. K. P. Zia, Phys. Rev. Lett. {\bf 66},
357 (1991)
\bibitem{brayscal} A. J. Bray, Adv. Phys. {\bf 43}, 357 (1994)
\bibitem{conig} A. Coniglio and M. Zannetti, Euro. Phys. Lett. {\bf 10}, 575
(1989)
\bibitem{emott} C. L. Emmott and A. J. Bray, Phys. Rev. E {\bf 59}, 213 (1999)
\bibitem{hum} A. J. Bray and k. Humayun, Phys. Rev. Lett. {\bf 68}, 1559
(1992) 
\bibitem{leu} K. -t. Leung and J. L. Cardy, J. Stat. Phys. {\bf 44}, 
567 (1986)
\bibitem{msr} P. C. Martin, E.D. Siggia and H. A. Rose, Phys. Rev. A {\bf 8}
, 423 (1973)
\bibitem{aha} A. Aharony, Phys. Rev. B {\bf 8}, 3363 (1973)
\bibitem{zinn} J. Zinn-Justin, {\it Quantum Field Theory 
and Critical Phenomena}, International series of monographs on Physics, 
Clarendon Press, Oxford, 3rd Edition, 1996

\end{references}
\end{document}